\RequirePackage{fix-cm}
\documentclass[twocolumn]{svjour3}          
\smartqed  
\usepackage{graphicx}
\usepackage{epstopdf}
\usepackage{amsmath}
\usepackage{algorithmic,algorithm}
\begin{document}

\title{Distributed Spectrum and Power Allocation for D2D-U Networks: A Scheme based on NN and Federated Learning\thanks{The article is an extended version of MONAMI 2020 conference
paper \cite{20_Zou}}
}


\author{Rui Yin  \and
        Zhiqun Zou \and
        Celimuge Wu \and
        Jiantao Yuan \and
        Xianfu Chen
}


\institute{Rui Yin \at
              School of Information and Electrical Engineering, Zhejiang University City College, Hangzhou 310015, China \\
              \email{yinrui@zucc.edu.cn}
           \and
           Zhiqun Zou \at
              College of Information Science and Electrical Engineering, Zhejiang University, Hangzhou, China
           \and
           Celimuge Wu (Corresponding author) \at
              Graduate School of Informatics and Engineering, The University of Electro-Communications,1-5-1, Chofugaoka, Chofu-shi, Tokyo 182-8585, Japan
           \and
           Jiantao Yuan \at
              Institute of Ocean Sensing and Networking of the Ocean College, Zhejiang University, Hangzhou, China
           \and
           Xianfu Chen \at
              VTT Technical Research Centre of Finland, Finland
}

\date{Received: date / Accepted: date}

\maketitle

\begin{abstract}
In this paper, a \emph{Device-to-Device communication on unlicensed bands} (D2D-U) enabled network is studied.
To improve the \emph{spectrum efficiency} (SE) on the unlicensed bands and fit its distributed structure while ensuring the fairness among D2D-U links and the harmonious coexistence with WiFi networks, a distributed joint power and spectrum scheme is proposed. In particular,
a parameter, named as price, is defined, which is updated at each D2D-U pair by a online trained \emph{Neural network} (NN) according to the channel state and traffic load.
In addition, the parameters used in the NN are updated by two ways, unsupervised self-iteration and federated learning, to guarantee the fairness and harmonious coexistence.
Then, a non-convex optimization problem with respect to the spectrum and power is formulated and solved on each D2D-U link to maximize its own data rate.
Numerical simulation results are demonstrated to verify the effectiveness of the proposed scheme.
\end{abstract}

\section{Introduction}
\label{intro}
The large-scale commercialization of the \emph{fifth generation} (5G) mobile
networks has brought us better communication experience with low latency and
high data rate
As a key technology in 5G networks, \emph{device-to-device} (D2D) communication allows direct transmission between D2D terminals instead of relaying through the base stations, which improves both system \emph{spectrum efficiency} (SE),
\emph{energy efficiency} (EE) and \emph{quality-of-service} (QoS) of D2D pairs \cite{09_Doppler}.

The conventional D2D communication mainly reuses the licensed channels with \emph{long-term evolution} (LTE) cellular networks to increase system SE and EE in the licensed spectrum \cite{14_Yu}.
However, the licensed spectrum is basically managed by mobile communication operators and expensive. In addition, with the explosive growth of the number of smart terminals, the spectrum resources on the licensed bands are becoming more scarce and D2D communications may cause severe interference to the original cellular users.
In order to guarantee the transmission performance of the cellular users as well as improve the QoS of D2D users, \emph{Device-to-Device on unlicensed bands} (D2D-U) is proposed to enable D2D communication on unlicensed spectrum \cite{16_Wu}. As the spectrum resources of unlicensed bands are abundant and free to use, D2D-U may significantly increase the SE and EE of D2D systems while guaranteeing the QoS of cellular
users \cite{16_Liu}.

Most existing works have studied on the mode selection, power and spectrum allocation mechanisms for D2D enabled cellular networks. In \cite{19_sha}, the impact of mode selection on effective capacity has been investigated via the Markov service process model.
Authors in \cite{14_xu} have proposed a centralized optimal mode selection and resource allocation for D2D enabled cellular networks. A distributed joint spectrum sharing and power allocation problem has been modeled as a non-convex optimization problem in \cite{16_Yin} and the suboptimal solution is obtained by convex approximation techniques. Similar problem is solved by a price-based model in \cite{15_Yin}, where a game-theoretic approach is proposed to mitigate interference among D2D pairs in a distributed way. Many machine learning-based methods have also been used to solve related problems in recent years. The authors of \cite{19_Lee} have designed a transmit power control strategy to D2D pairs based on a \emph{deep neural network} (DNN) structure, where the SE and interference are taken into account. A deep reinforcement learning-based method is utilized in \cite{18_Mou} to maximize the sum rates of D2D links.

Recently, \emph{long-term evolution on unlicensed bands} (LTE-U) system is introduced into the unlicensed spectrum. \emph{listen-before-talk} (LBT) and \emph{duty cycle method} (DCM) access mechanisms have been proposed for LTE-U based cellular users to access the unlicensed spectrum while ensuring the fair coexistence with WiFi networks \cite{lbtdcm1,lbtdcm2,lbtdcm3,20_sun,20_cui}. In \cite{16_yin2}, the back-off window size based on LBT mechanism is adaptively adjusted according to the WiFi traffic load and available bandwidth on licensed spectrum, which improves the system spectrum efficiency. A Q-learning based scheme has been also proposed to adjust the back off window size of LBT in \cite{18_Vasilis}. The performance of DCM mechanism has been analyzed in \cite{19_Santana,20_neto}, where the reinforcement learning based methods have been employed to achieve fair coexistence between LTE-U and WiFi networks. A hybrid mechanism has been designed in \cite{19_Liu}, both LBT and DCM have been utilized and the flexible handoff between two mechanisms is achieved to meet fairness constraint.

Only a small amount of work has focused on D2D transmission on unlicensed bands. The conclusion of \cite{15_andr} has proven that D2D-U can significantly mitigate the congestion, conflicts and improve the system throughput.
In \cite{17_Zhang}, the sub-channel allocation of D2D-U enabled LTE cellular networks has been formulated as a many-to-many matching problem with externality and an iterative sub-channel swap algorithm has been proposed to improve the system performance.
A reinforcement learning based scheme is proposed in \cite{19_Zou}, where a deep Q-network has been utilized to
learn the traffic load on the unlicensed spectrum. It allows D2D-U system to model the joint allocation problem as a convex optimization problem. A decentralized joint spectrum and power allocation approach for D2D-U has been proposed in \cite{20_Yin}, which can meet the global minimization of power consumption among the D2D-U pairs.

After thorough investigation, most above mentioned power and spectrum allocation schemes are centralized, which may bring large signaling overhead to the base station and lead to high latency.
Besides, most of the work concentrates only on maximizing system throughput without considering the different traffic loads of different D2D pairs.
In this paper, we first define the unlicensed channel traffic load according to the number of competing WiFi users when DCM scheme is applied at D2D-U network.
A price based model where D2D-U users need to pay for the channel resources is then built and a \emph{Neural network} (NN) is applied to estimate the price to use unlicensed spectrum at each D2D-U link adaptively.
In order to guarantee the harmonious coexistence with WiFi networks and the fairness among D2D-U pairs, the loss function is designed specifically to realize the online unsupervised learning of NN.

However, the training of NN in a distributed system is always unstable. If individual D2D-U links experience excessive noise or over fitting problem, it will cause severe interference to the performance of the whole system. In addition, when new D2D-U links join the network, the randomly initialized NN parameters will also cause serious fluctuations to the system which has converged.
To solve above problem, the federated learning method \cite{17_Kon} is utilized to update the parameters of the networks. In the scenario of federated learning, distributed terminals train NNs themselves, and then integrate gradients or parameters at the center base station periodically \cite{19_Nik}. In our model, the integrated parameters can help correct those networks with large deviations and also initialize parameters for new D2D-U users.
Afterwards, with the corresponding prices of channels, the spectrum and power allocation can be optimized jointly to maximize the transmission rates at each D2D-U pair.

The main contributions of the paper are summarized as follows.
\begin{enumerate}
  \item  A DCM based channel access model is built for D2D-U networks to share the unlicensed spectrum with WiFi networks. According to the \emph{carrier sensing multiple access with collision avoidance} (CSMA/CA) mechanism adopted in WiFi, a novel traffic load on the unlicensed channel is defined.

  \item To balance the SE and traffic load of D2D-U pairs while ensuring the fairness, a virtual variable, named as price, is defined,
      which is related to the traffic load, channel state of D2D-U links and the total traffic load on the unlicensed channels. With the price, the transmission power and unlicensed spectrum allocation can be optimized jointly in a distributed way at each D2D-U pair.

  \item Since it is hard to formulate an explicit function to model the relationship between the price, the traffic load and channel state information, an online trained NN with a specific loss function is applied to update the price at each D2D-U pair adaptively. The parameters of NN are updated via unsupervised self-iteration as well as federated learning to stabilize the system performance.

  \item The centralized optimal solution is presented for comparison. Moreover, the simulation results are provided to verify the effectiveness of the scheme and the theoretical analysis.

\end{enumerate}

The rest of this paper is organized as follows. Section~\ref{s2} introduces the
system model and a novel definition of WiFi traffic load on unlicensed channels.
The price based learning model is proposed in Section~\ref{s3} and Section~\ref{s4}, respectively.
We analyze the simulation results in Section~\ref{s5} and summarize the paper in Section~\ref{s6}

\section{System Model}\label{s2}
\begin{figure}

\includegraphics[width=0.5\textwidth]{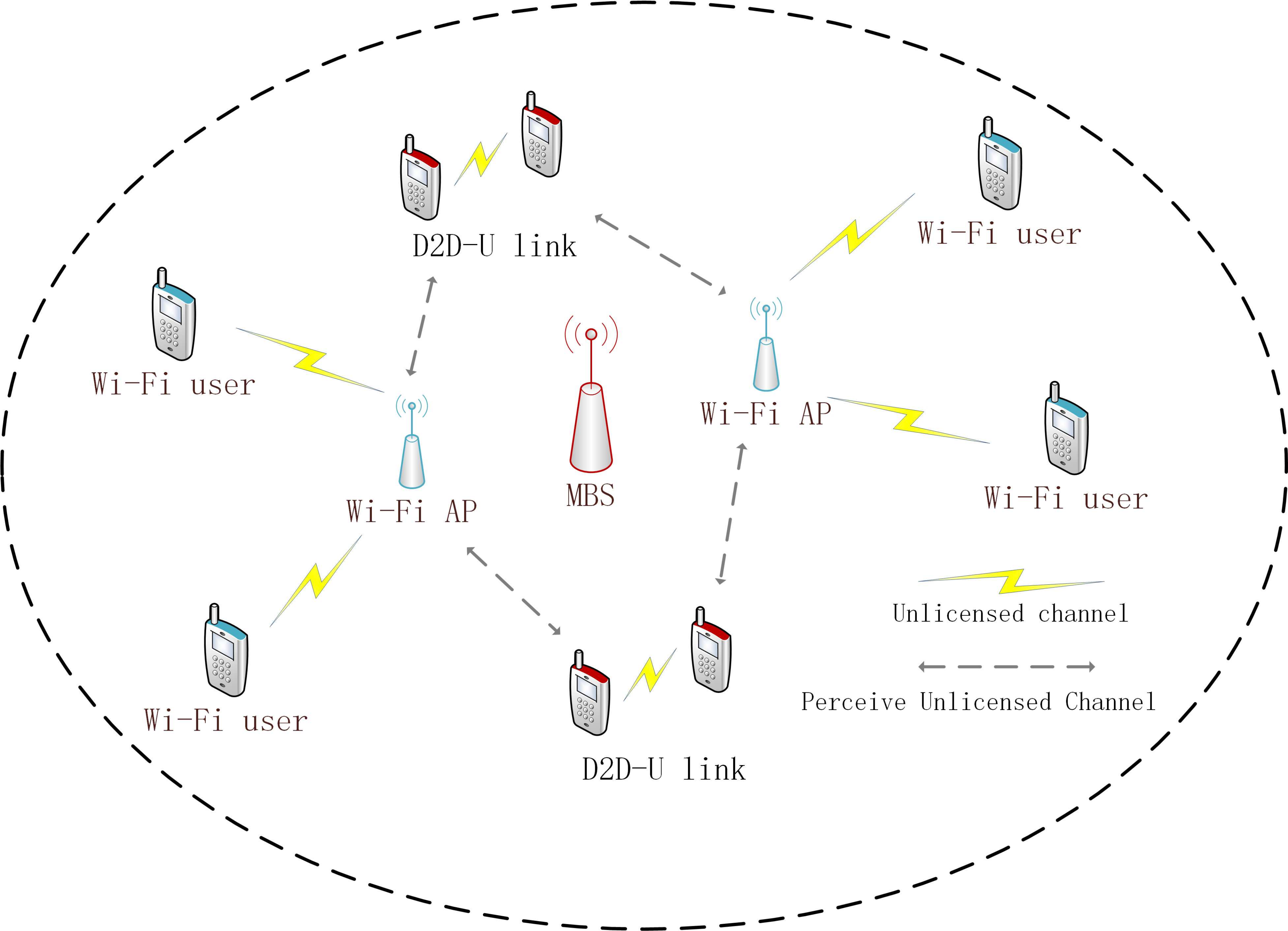}
\caption{System model.} \label{fig1}
\end{figure}

In this paper, we study the scenario where multiple D2D-U links share the unlicensed channels with WiFi \emph{Access points} (APs), as shown in Fig.~\ref{fig1}, where D2D-U links are able to simultaneously use multiple unlicensed channels and a single unlicensed channel can be shared by more than one D2D-U pair.
\emph{Macro base station} (MBS) can obtain the information on the achievable data rates of each D2D-U pair and the parameters of NNs via the control channel on licensed bands.
It is worth mentioning that, in order to consider more realistic scenarios, the number of D2D-U links is not fixed, which means that D2D-U pairs may leave and join the network dynamically.
To model the system mathematically, we use set $\mathcal{D}=\{d_0,d_1,...,d_{N-1}\}$ to demonstrate $N$ D2D-U links in the coverage of the MBS and $N$ is not fixed. Moreover, there are $M$ accessible unlicensed channels, denoted by set, $\mathcal{U}$, $\mathcal{U}=\{u_0,u_1,...,u_{M-1}\}$, which are orthogonal with each other. To consider the fairness among D2D-U links and the harmonious coexistence with WiFi networks, $\mathcal{L^D}=\{l^D_0,l^D_1,...,l^D_{N-1}\}$ and $\mathcal{L^U}=\{l^U_0,l^U_1,...,l^U_{M-1}\}$ are used to denote the traffic loads of D2D-U links and WiFi systems on the unlicensed channels, respectively.
In addition, WiFi APs adopt the CSMA/CA based \emph{distributed coordination function} (DCF) to
access the unlicensed channels while DCM mechanism is applied at D2D-U links to access the unlicensed channels.

\subsection{Achievable data rates at D2D-U links}
\begin{figure}
\includegraphics[width=0.5\textwidth]{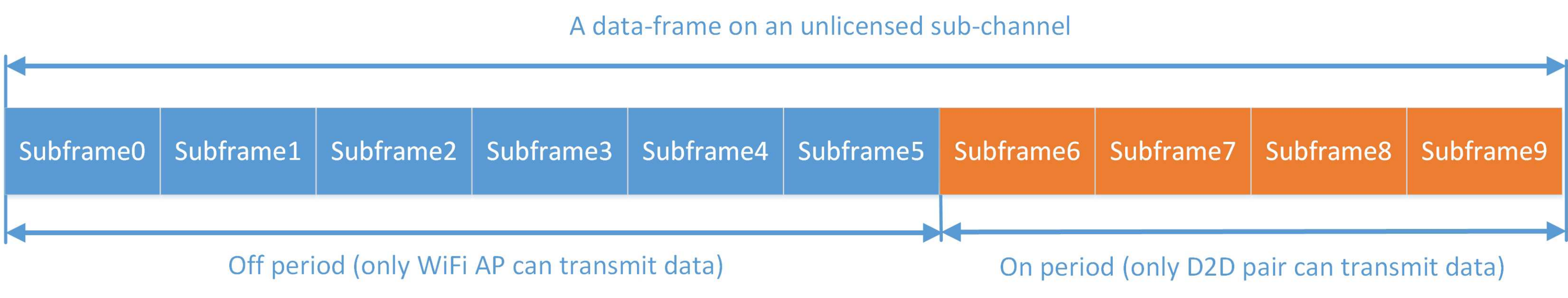}
\caption{DCM mechanism.} \label{fig2}

\end{figure}

The transmission mode when applying DCM mechanism at D2D-U links is shown in Fig.~\ref{fig2}, where the time frames on the unlicensed channels are divided into two parts.
WiFi users can only transmit data during the first part, which is called `off period', and the remaining section, named as `on period', is occupied by D2D-U pairs. We further use $\theta_{i,j}$, $\theta_{i,j} \in [0,1]$, to represent the proportion used by D2D-U link $d_i$ on unlicensed channel $u_j$. Then, the achievable data-rate at $d_i$ on unlicensed channels can be calculated by
\begin{equation}\label{datarate_D2DU}
R_{i} = \sum_{j=0}^{M-1}\theta_{i,j}B_j\log\left(1+\frac{p_{i,j}h_{i,j}}{N_0B_j}\right),
\end{equation}
where $B_j$ is the bandwidth of unlicensed channel $u_j$, $h_{i,j}$ and $p_{i,j}$ are the channel power gain and corresponding transmission power of D2D-U pair $d_i$ on $u_j$, respectively. $N_0$ is the noise power spectrum on unlicensed channel, which is fixed in the manuscript.

\subsection{WiFi traffic load definition}
To ensure the transmission requirements of WiFi users, D2D-U links must decide the duration of `on period' based on the WiFi traffic load on the unlicensed channels, which means that D2D-U links need to obtain WiFi traffic load before accessing the channels.
The conventional traffic load of WiFi networks is mainly decided by the number of competing WiFi users. As demonstrated in \cite{03_Bianchi,13_Qin}, the extended Kalman filter has been used to achieve an accurate estimation on the number of active WiFi users, where D2D-U links filter the transmission collision probability in the channel and then obtain the number of WiFi users based on the estimated transmission collision probability.
However, the impact of the number of WiFi users on the throughput of the WiFi system is non-linear and D2D-U links cannot directly determine the duration of `on period' based on the number of WiFi users. Therefore, a novel WiFi traffic load definition is first defined when the DCM mechanism is employed at the D2D-U pairs.

\begin{figure}

\includegraphics[width=0.5\textwidth]{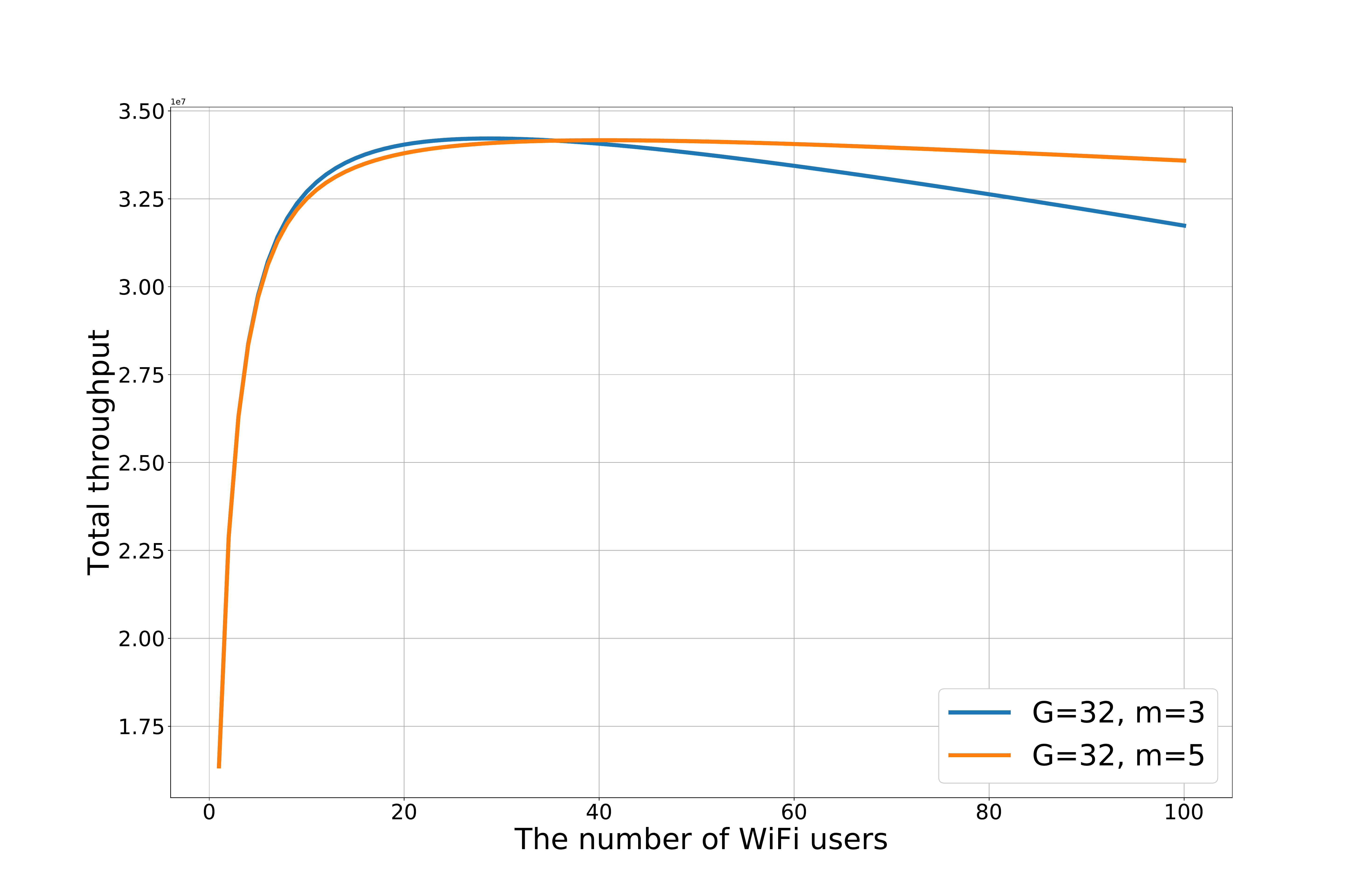}
\caption{Relationship between throughput and number of WiFi users.} \label{fig3}

\end{figure}

As WiFi APs adopt binary slotted exponential back-off scheme in DCF, the relationship between the total WiFi throughput on an unlicensed channel and the number of WiFi users could be obtained according to \cite{00_Bianchi}, as illustrated in Fig~\ref{fig3}. The size of back-off contention window, denoted as $G$, is $32$ and the maximum back-off contention stage, denoted as $m$, is set to $3 $ and $5$, respectively. We can observe that, as the number of WiFi users increases, the achievable throughput on the unlicensed channel increases first and then decreases.
The reason is that when a large number of WiFi users compete for the same unlicensed channel, the transmission collision probability will increase, resulting in transmission failure and lower throughput.

Herein, we use the number of WiFi users corresponding to the highest throughput, $n_k^{\max}$, to represent the maximum load that the WiFi network can handle on unlicensed channel $u_k$. If the number of WiFi users in the unlicensed channel is greater or equal to $n^{\max}_k$, the channel $u_k$ is considered inaccessible to D2D-U pairs.
Furthermore, in order to define the WiFi traffic load to fit the DCM mechanism, when the throughput of the WiFi network reaches maximum, the average throughput of each WiFi user, $\hat r_{k}^{\max}$, is treated as the basic throughput guarantee of WiFi users.
Let $R^{\max}_{k}$ denote the maximum system throughput on unlicensed channel $u_k$. Then, $\hat r^{\max}_{k}$ can be calculated as $\hat r_{k}^{\max} = \frac{R_{k}^{\max}}{n_k^{\max}}$. The basic throughput guarantee means that when D2D-U pairs reuse the unlicensed channels based on DCM mechanism, the average throughput of WiFi users should not be less than $\hat r^{\max}_{k}$.

\begin{figure}
\includegraphics[width=0.5\textwidth]{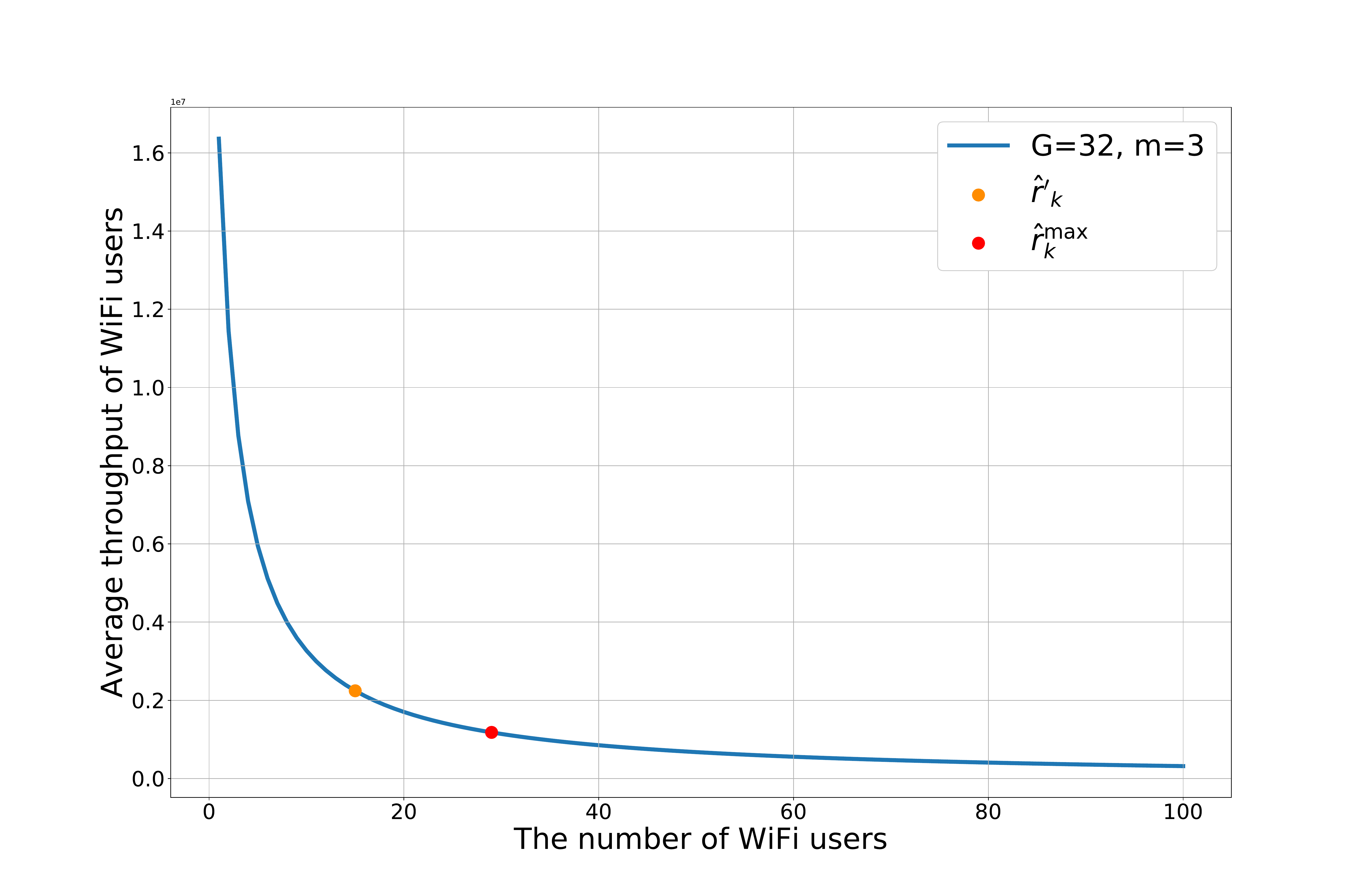}
\caption{Relationship between $\hat r_k'$ and $\hat r^{\max}_{k}$.} \label{avr}
\end{figure}

Then we can calculate the minimum value of the `off period' based on the above description. For unlicensed channel $u_k \in \mathcal{U}$, let $\hat r_k$ represent the average throughput of each WiFi user when no D2D-U links use $u_k$. On the other hand, when the number of WiFi users is less than $n^{\max}_k$, D2D-U pairs are allowed to use $u_k$ with DCM mechanism and the average throughput of WiFi users is given by
\begin{equation}\label{rk}
\hat r_k' = \hat r_k(1 - \sum_{i = 0}^{N - 1}\theta_{i,k}).
\end{equation}
According to the basic throughput guarantee, we can further achieve
\begin{equation}\label{rk2}
\hat r^{\max}_{k} = \frac{R^{\max}_{k}}{n^{\max}_k} \le \hat r_k'.
\end{equation}
The relation ship between $\hat r_k'$ and $\hat r^{\max}_{k}$ is shown in Fig~\ref{avr}, when $\hat r_k'$ locates on the left side of $\hat r^{\max}_{k}$, $u_k$ is available to D2D-U users.
In order to adapt to DCM access mechanism, the WiFi traffic load $l^U_k$ on $u_k$ is defined as the minimum `off period' duration that meets the basic throughput guarantee of WiFi users. Combining \eqref{rk} and \eqref{rk2}, $l^U_k$ can be given by
\begin{equation}\label{luk}
l^U_k = \frac{\hat r^{\max}_{k}}{\hat r_k} \le 1 - \sum_{i = 0}^{N - 1}\theta_{i,k}.
\end{equation}

Since both $\hat r^{\max}_{k}$ and $\hat r_k$ can be calculated according to the physical layer parameters of the WiFi networks \cite{00_Bianchi} and the number of WiFi users can be accurately estimated by extended Kalman filter based methods \cite{03_Bianchi,13_Qin} and learning based method \cite{20_Yin2}, D2D-U links are able to sense the traffic load in the unlicensed channel and then decide their own resource allocation scheme. In next section, the resource allocation model is built for D2D-U links and a priced-based solution is proposed.

\section{Distributed price based Model}\label{s3}
In this section, we first formulate a distributed optimization problem for each D2D-U link to maximize its own data rate.
Then, in order to ensure fairness among D2D links, a priced-based solution is proposed to provide D2D-U links with different prices for using unlicensed spectrum under different traffic load and channel state conditions.

\subsection{Problem formulation}
For a single D2D link, $d_i \in \mathcal{D}$, to maximize its transmission rates while guaranteeing the performance of WiFi networks, an optimization problem can be formulated as
\begin{equation}\label{op}
\max_{\{\theta_{i,j}, p_{i,j}\}}{\{R_{i}\}},
\end{equation}
subject to
\begin{align}
&C1: \theta_{i,j} \le 1 - l_j^{U},~\forall j\in [0, M-1], \label{1_c1}\tag{\theequation a}\\
&C2: \sum_{j = 0}^{M-1} \theta_{i,j}p_{i,j} \le p_c, \label{1_c2}\tag{\theequation b}\\
&C3: \theta_{i,j}p_{i,j} \le p_u,~\forall j\in [0, M-1],\label{1_c3}\tag{\theequation c}
\end{align}
where $C1$ is to guarantee the fair coexistence with WiFi networks on the unlicensed channels, $C2$ is the total power constraint of $d_i$ and $C3$ is the transmission power limit on the unlicensed channel according to the regulation.

Problem \eqref{op} is a non-convex problem but can be converted into a convex optimization problem and solved on each D2D link. In detail, a extra parameter $\eta_{i,j} = \theta_{i,j}p_{i,j}$ is introduced and \eqref{datarate_D2DU} can be re-expressed as
\begin{equation}\label{datarate_D2DU2}
R_{i} = \sum_{j=0}^{M-1}\theta_{i,j}B_j\log\left(1+\frac{\eta_{i,j}h_{i,j}}{N_0B_j\theta_{i,j}}\right),
\end{equation}
then problem \eqref{op} is converted into
\begin{equation}\label{op12}
\max_{\{\theta_{i,j}, \eta_{i,j}\}}{\{R_{i}\}},
\end{equation}
subject to
\begin{align}
&C1: \theta_{i,j} \le 1 - l_j^{U},~\forall j\in [0, M-1], \label{1_c1}\tag{\theequation a}\\
&C2: \sum_{j = 0}^{M-1} \eta_{i,j} \le p_c, \label{1_c2}\tag{\theequation b}\\
&C3: \eta_{i,j} \le p_u,~\forall j\in [0, M-1].\label{1_c3}\tag{\theequation c}
\end{align}
Problem \eqref{op12} is a convex optimization problem and can be solved by Lagrangian multiplier method.
However, optimization problem \eqref{op12} can only allow a D2D-U link to maximize its own throughput under the constraint of guaranteeing the fair coexistence with WiFi networks without considering the impact on other D2D-U pairs.
When multiple D2D-U links share the same unlicensed channel, the possibility of transmission collision is extremely high, which leads to the
lose on the performance of the D2D-U transmission.
Therefore, the model needs to be improved based on the respective traffic load conditions of D2D-U links, where D2D-U links with heavy transmission tasks could use more spectrum resources while D2D-U links with light transmission tasks require only a small fraction of unlicensed spectrum resources. In next subsection, a priced-based solution is applied to achieve this goal.

\subsection{Priced-based solution}
Different from the Stackelberg game based method in \cite{15_Yin,20_Zhou,20_Gu} to minimize users' overhead, our method adjusts the price when assets are limited.
In the proposed price-based model, each D2D-U link is considered as a consumer and spectrum resources in unlicensed bands are provided to consumers as commodities.
The total money which each D2D-U link has are set to the same, which is represented by $C$. Define the price corresponding to the unlicensed channel $u_j$ for the D2D-U link $d_i$ as $c_{i,j}$ and when $d_i$ transmits data on $u_j$, the price $d_i$ needs to pay is written as $\theta_{i,j}\times c_{i,j}$.
Accordingly, the optimization problem(\ref{op}) can be expressed with an extra fairness constraint as
\begin{equation}\label{op2}
\max_{\{\theta_{i,j}, p_{i,j}\}}{\{R_{i}\}},
\end{equation}
subject to
\begin{align}
&C1: \theta_{i,j} \le 1 - l_j,~\forall j\in [0, M-1], \label{1_c1}\tag{\theequation a}\\
&C2: \sum_{j = 0}^{M-1} \theta_{i,j}p_{i,j} \le p_c, \label{1_c2}\tag{\theequation b}\\
&C3: \theta_{i,j}p_{i,j} \le p_u,~\forall j\in [0, M-1],\label{1_c3}\tag{\theequation c}\\
&C4: \sum_{j=0}^{M-1} \theta_{i,j}c_{i,j} \le C. \label{1_c4}\tag{\theequation d}
\end{align}
The above problem is also a non-convex optimization problem and can not be solved in its current formation. Same as in Problem \eqref{op}, replacing $p_{i,j}$ with $p_{i,j}=\frac{\eta_{i,j}}{\theta_{i,j}} $ and the above problem is converted into
\begin{equation}\label{op22}
\max_{\{\theta_{i,j}, \eta_{i,j}\}}{\{R_{i}\}},
\end{equation}
subject to
\begin{align}
&C1: \theta_{i,j} \le 1 - l_j,~\forall j\in [0, M-1], \label{1_c1}\tag{\theequation a}\\
&C2: \sum_{j = 0}^{M-1} \eta_{i,j} \le p_c, \label{1_c2}\tag{\theequation b}\\
&C3: \eta_{i,j} \le p_u,~\forall j\in [0, M-1],\label{1_c3}\tag{\theequation c}\\
&C4: \sum_{j=0}^{M-1} \theta_{i,j}c_{i,j} \le C. \label{1_c4}\tag{\theequation d}
\end{align}
Herein, one important key to solve \ref{op22} is to find $c_{i,j}$. If $c_{i,j}$ is known, the above problem is a convex optimization problem and the optimal solution can be obtained based on the Lagrangian multiplier method.
The Lagrangian function of Problem \ref{op22} is constructed as
\begin{equation}\label{lag1}
\begin{aligned}
&L(\theta_{i,j},\eta_{i,j}, \mu_j^{(1)},\mu^{(2)}, \mu_j^{(3)}, \mu^{(4)})=-R_i \\
&+ \sum_{j=0}^{M-1}\mu_j^{(1)}(\theta_{i,j}+l_j-1) + \mu^{(2)}(\sum_{j=0}^{M-1} \eta_{i,j} - C)\\
&+\sum_{j=0}^{M-1}\mu_j^{(3)}(\eta_{i,j} - p_u)+ \mu^{(4)}(\sum_{j=0}^{M-1}\theta_{i,j}c_{i,j}-C),
\end{aligned}
\end{equation}
where $\mu_j^{(1)}$, $\mu^{(2)}$, $\mu_j^{(3)}~{\rm{and}}~\mu^{(4)}$ are the Lagrangian multipliers and the \emph{Karush-Kuhn-Tucker} (KKT) conditions of $L(\cdot)$ are derived based on \eqref{lag1} as
\begin{equation}\label{KKT1_1}
\frac{\partial L}{\partial \theta_{i,j}} = 0, ~\forall j\in [0, M-1],
\end{equation}
\begin{equation}\label{KKT1_2}
\frac{\partial L}{\partial \eta_{i,j}} = 0, ~\forall j\in [0, M-1],
\end{equation}
\begin{equation}\label{KKT1_3}
\mu_j^{(1)}(\theta_{i,j}+l_j-1) = 0, ~\forall j\in [0, M-1],
\end{equation}
\begin{equation}\label{KKT1_4}
\mu^{(2)}(\sum_{j=0}^{M-1} \eta_{i,j} - C) = 0,
\end{equation}
\begin{equation}\label{KKT1_5}
\mu_j^{(3)}(\eta_{i,j} - p_u) = 0, ~\forall j\in [0, M-1],
\end{equation}
\begin{equation}\label{KKT1_6}
\mu^{(4)}(\sum_{j=0}^{M-1}\theta_{i,j}c_{i,j}-C) = 0,
\end{equation}
\begin{equation}\label{KKT1_7}
\mu_j^{(1)} \ge 0, \mu^{(2)} \ge 0, \mu_j^{(3)} \ge 0, \mu^{(4)} \ge 0, ~\forall j\in [0, M-1].
\end{equation}
On the basis of KKT conditions, the optimal solutions of $\theta_{i,j}$ and $\eta_{i,j}$ should satisfy
\begin{equation}\label{KKT1_8}
\eta_{i,j} = \theta_{i,j}B_{j}(\frac{\log e}{\mu^{(2)} + \mu_j^{(3)}} -\frac{N_0}{h_{i,j}}),
\end{equation}
and
\begin{equation}\label{KKT1_9}
\log(1+\frac{\eta_{i,j}h_{i,j}}{N_0B_j\theta_{i,j}}) - \frac{\eta_{i,j}h_{i,j}\log e}{N_0B_j\theta_{i,j}+\eta_{i,j}h_{i,j}} = \frac{\mu_j^{(1)}+\mu^{(4)}c_{i,j}}{B_j}.
\end{equation}
Then, according to \eqref{KKT1_8} and \eqref{KKT1_9}, $\eta_{i,j}$ and $\theta_{i,j}$
can be achieved for different $d_i$ and $u_j$. Since \eqref{KKT1_9} is a
transcendental equation, numerical method can be applied to find the solution.

Based on above analysis, $\theta_{i,j}$ and $p_{i,j}$ can be obtained with the known price, $c_{i,j}$, which can be used to ensure the fairness among D2D-U pairs. To adjust prices adaptively to reach the fairness, each D2D-U pair needs to
determine the $c_{i,j}$ based on its own traffic load, channel state information and the WiFi traffic load on the unlicensed channel. Accordingly, we denote $c_{i,j}$ as
\begin{equation}\label{cij}
c_{i,j} = F(l^D_i,l^U_j,h_{i,j}|s^c),
\end{equation}
where $l^D_i$ represents the transmission task of $d_i$, $h_{i,j}$ represents the channel power gain on unlicensed channel $u_j$. The function $F(\cdot)$ is to model the relationship between the price and the traffic loads of the D2D links and the unlicensed channels.  When the traffic load of D2D-U link, $d_i$, is heavy, $d_i$ is encouraged to use unlicensed spectrum with a low price while the price to D2D-U link with less traffic load is high; Moreover, unlicensed channels with low traffic loads or strong channel power gain for the D2D link will need to be paid with low prices while under channels with high traffic loads or poor channel conditions will be expensive.
In addition, to mitigate the channel access conflict among D2D-U links, a feedback signal $s^c$ is set on $d_i$. If $d_i$ collides with other D2D-U links on the channel, $s^c$ is activated and the price should be enhanced accordingly.
Therefore, function  $F(\cdot)$ should have the following characteristics:
\begin{itemize}
  \item [(1).] $F(\cdot)$ should decrease monotonically with respect to $l^D_i$;
  \item [(2).] $F(\cdot)$ should increase monotonically with respect to $l^U_j$;
  \item [(3).] $F(\cdot)$ should decrease monotonically with respect to $h_{i,j}$;
  \item [(4).] $F(\cdot)$ should increase with the activation of $s^c$.
\end{itemize}

As for the fairness among D2D-U links, we define \emph{Expected transmission time} (ETT) as
the ratio of a D2D-U link's traffic load to its achievable data rates. Then, the fairness sharing
on unlicensed channels among D2D-U links is denoted as that the ETT values of all D2D-U links are equal, which is written as
\begin{equation}\label{fair_tr}
\frac{l^D_0}{R_0} = \frac{l^D_1}{R_1} = \cdots = \frac{l^D_{M-1}}{R_{M-1}}.
\end{equation}

However, it is difficult to directly build an explicit mathematical model to formulate the function $F(\cdot)$ and achieve \eqref{fair_tr}. To address this issue, an online training \emph{neural network} (NN) architecture is exploited to implement function of $F(\cdot)$ and the loss function for all D2D-U links are provided based on $s^c$ and the assistance of MBS. Specific details of the adopted
NN will be given in the next section.

\section{Learning based Method}\label{s4}

\begin{figure}

\includegraphics[width=0.5\textwidth]{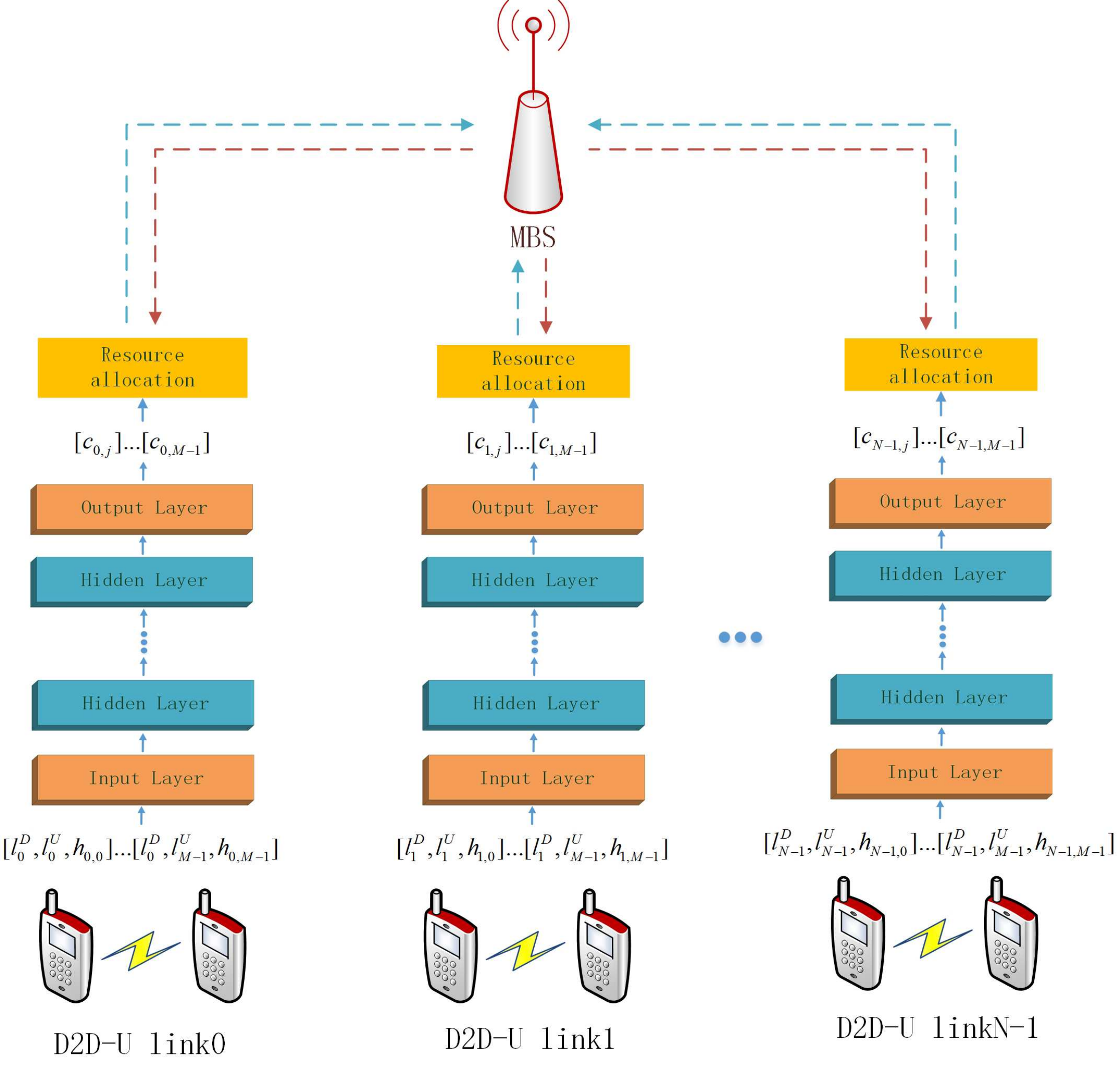}
\caption{The structure of NN.}\label{NNfig}

\end{figure}

Because of the strong fitting performance and robustness of NN, the online trained NN is utilized to achieve adaptive adjustment of the prices in a dynamic D2D-U environment, where the parameters of NN are updated according to the loss function as well as the federated learning.
The structure of the whole distributed price system is illustrated in Fig.~\ref{NNfig} where each D2D-U pair holds a NN itself and determines its own prices of different channels.
In this section, the structure and the loss function of NN will be first introduced to train NN in an unsupervised online mode, then to improve the stability of the system and better accept new D2D-U users, the federated learning based mode will be proposed to adjust parameters in the networks.

\subsection{Unsupervised online training method}
As demonstrated in Fig.~\ref{NNfig}, the output of NN in D2D-U link $d_i$ can be calculate as
\begin{equation}\label{NN}
\hat c_{i,j} = \hat F(l^D_i,l^U_j,h_{i,j}|\phi),
\end{equation}
where $\hat c_{i,j}$ is the output as well as the price estimated by NN, the input of NN is $[l^D_i,l^U_j,h_{i,j}]$.
In particular, before $[l^D_i,l^U_j,h_{i,j}]$ is feeded into NN, the input data needs to be normalized to avoid problems caused by different orders of magnitude.
$\hat F(\cdot)$ indicates the forward propagation of NN and $\phi$ represents all the weights and bias parameters.
At each iteration, all the parameters in NN are updated based on the gradient descent algorithm, which is denoted by
\begin{equation}\label{NN_TRAIN}
\phi = \phi - \alpha \frac{\partial Q}{\partial \phi},
\end{equation}
where $\alpha$ is the learning rate of NN and $Q$ is the loss function. Since it is hard to achieve global optimal solution of problem \eqref{op2}, we cannot obtain labels and use the supervised learning method to train the network. Therefore, based on \eqref{fair_tr} and the collision detection, the loss function $Q$ is formulated by two parts to train NN in an unsupervised way.
$Q_1$ and $Q_2$ are used to represent these two parts, respectively, where $Q_1$ is defined as:
\begin{itemize}
  \item [(1).] if $\frac{l^D_i}{R_i}$ is larger than ETT values of $\frac{M+1}{2}$ D2D-U links (when $M$ is odd) or $\frac{M}{2}$ D2D-U links (when $M$ is even), $Q_1 = q$;
  \item [(2).] if $\frac{l^D_i}{R_i}$ is smaller than ETT values of $\frac{M+1}{2}$ D2D-U links (when $M$ is odd) or $\frac{M}{2}$ D2D-U links (when $M$ is even), $Q_1 = -q$;
  \item [(3).] else $Q_1 = 0$;
\end{itemize}
where $q$ is the adjustment step size of the price and is set to a tiny positive value.
$Q_2$ corresponds to conflict feedback $s^c$, which is defined as:
\begin{itemize}
  \item [(1).] if $d_i$ collides with other D2D-U links, $Q_2$ is set to $v_1$;
  \item [(2).] else $Q_2 = v_2$;
\end{itemize}
In order to mitigate the transmission collision among D2D-U links effectively in actual operation, $v_1$ is set to a larger positive value to significantly increase the price of the unlicensed channel when collision happens.
$v_2$ is a negative value which aims at decreasing prices to allow D2D-U links to use more spectrum resources when no collision happens.
Herein, the value $Q_1$ can be provided by MBS and $Q_2$ can be decided on $d_i$ according to its transmission collision situation.
Accordingly, the target of NN output is denoted as $T = \hat c_{i,j} + Q_1 + Q_2$ and $Q$ can be calculated by Mean-Squared Loss as
\begin{equation}\label{loss}
Q = (T - \hat c_{i,j})^2 = (Q_1+Q_2)^2.
\end{equation}
More specifically, in each time slot, $d_i$ will obtain the prices of $M$ unlicensed channels and get $M$ loss values based on \eqref{loss}. Here $M$ training data will be treated as a batch to jointly train the network.
Furthermore, to keep the convergence of NN and the stability of output, we use Sigmoid function to limit the output value in a certain range according to the actual conditions. The activation function of the output layer is set to be $w\times \rm{Sigmoid}(\cdot)$, which limits
the output in $[0,w]$.

It is noteworthy that each D2D-U link holds a NN independently to determine the price corresponding to the utilization on the unlicensed channels. When the neural networks of D2D-U links converge, the system has reached an equilibrium. When the traffic load of D2D-U links or WiFi system changes, neural networks will converge to a new equilibrium adaptively.
However, in such an online dynamic environment, the network may be hard to converge due to environmental noise or over fitting. Due to the setting of the loss function, the un-convergence of a single NN may affect the updates of other D2D-U users, which leads to poor system performance. Additionally, when new D2D-U pairs join in the system, their NN parameters will be initialized randomly, which can also cause the instability of the system. To solved the mentioned problem, the federated learning based mode will be utilized to further adjust NN parameters.

\subsection{Federated learning based method}
Federated learning based schemes have achieved significant performance in distributed scenarios \cite{19_Yang}. Without a large amount of data interaction, federated learning ensures safety and stability of the system by integrating the gradient or parameters information of distributed users at the central processor and then feedback to the distributed terminals.
In our D2D-U model, each D2D-U pair is a distributed terminal and MBS plays the role as the center. Then MBS periodically collects NN parameters of D2D-U pairs in the region and averages them, which can be denoted as
\begin{equation}\label{av_phi}
\hat \phi = \frac{\sum_{i=0}^{N-1}\phi_i}{N},
\end{equation}
where $\phi_i$ is NN parameters of D2D-U link $d_i$. Since the traffic load and channel gain of different D2D-U links are not the same, the information contained in the neural network corresponding to the expected value of $\hat \phi$, $E(\hat \phi)$, is the average traffic load and transmission status of all D2D-U users in the region.

Then for new D2D-U pairs who join the system, instead of initializing NN parameters randomly, we use current $\hat \phi$ saved by MBS as the starting point of NN for training. It can effectively reduce the instability of the initial training of NN and accelerate the network convergence.
For existing D2D-U pair $d_i$, in each iteration of parameter updates based on federated learning, $d_i$ needs to adjust the value of $\phi_i$ according to $\hat \phi$. If NN in $d_i$ has converged and output stable price value, $\phi_i$ should be kept or update little. On the other hand, if the output is unstable or the loss value is large, $\phi_i$ need correction. The update of $\phi_i$ is calculated by
\begin{equation}\label{av_phi_phi_i}
\phi_i = \beta \hat \phi + (1-\beta)\phi_i,
\end{equation}
where $\beta \in (0,1)$ is the update intensity parameter which is determined adaptively by the network convergence.
In order to realize the judgment of network convergence, here the accumulated absolute loss value of NN is utilized to be the basis for judgment.
Let $Q_{sum,i}$ be the accumulated absolute loss value of $d_i$, if the value of $Q_{sum,i}$ is tiny, which means that NN has converged and $\beta$ should be small. When $Q_{sum,i}$ is a large number, $\beta$ need to be set large accordingly.
Here we use the sigmoid function again to express the relationship between $Q_{sum,i}$ and $\beta$ as
\begin{equation}\label{q_beta}
\beta = \frac{1}{1+e^{-(\frac{\gamma}{\epsilon} Q_{sum,i}-\gamma)}},
\end{equation}
where $\gamma$ and $\epsilon$ are constant value and decided by specific environment and the loss value. Figure.~\ref{q_beta} describes this relationship more intuitively, we can find that when $Q_{sum,i}$ is small, the value of $\beta$ is close to $0$, as $Q_{sum,i}$ increases, $\beta$ also increases till close to $1$.

Based on the above interpretation on the proposed NN and the federated learning based mode, the process of the joint power and spectrum allocation algorithm for a single D2D-U link, $d_i \in \mathcal{D}$, is summarized in Algorithm~\ref{algr}. $T$ is the system clock recorded by MBS and $T_{fl}$ is the length of the federated learning period.

\begin{figure}
\begin{centering}
\includegraphics[width=0.5\textwidth]{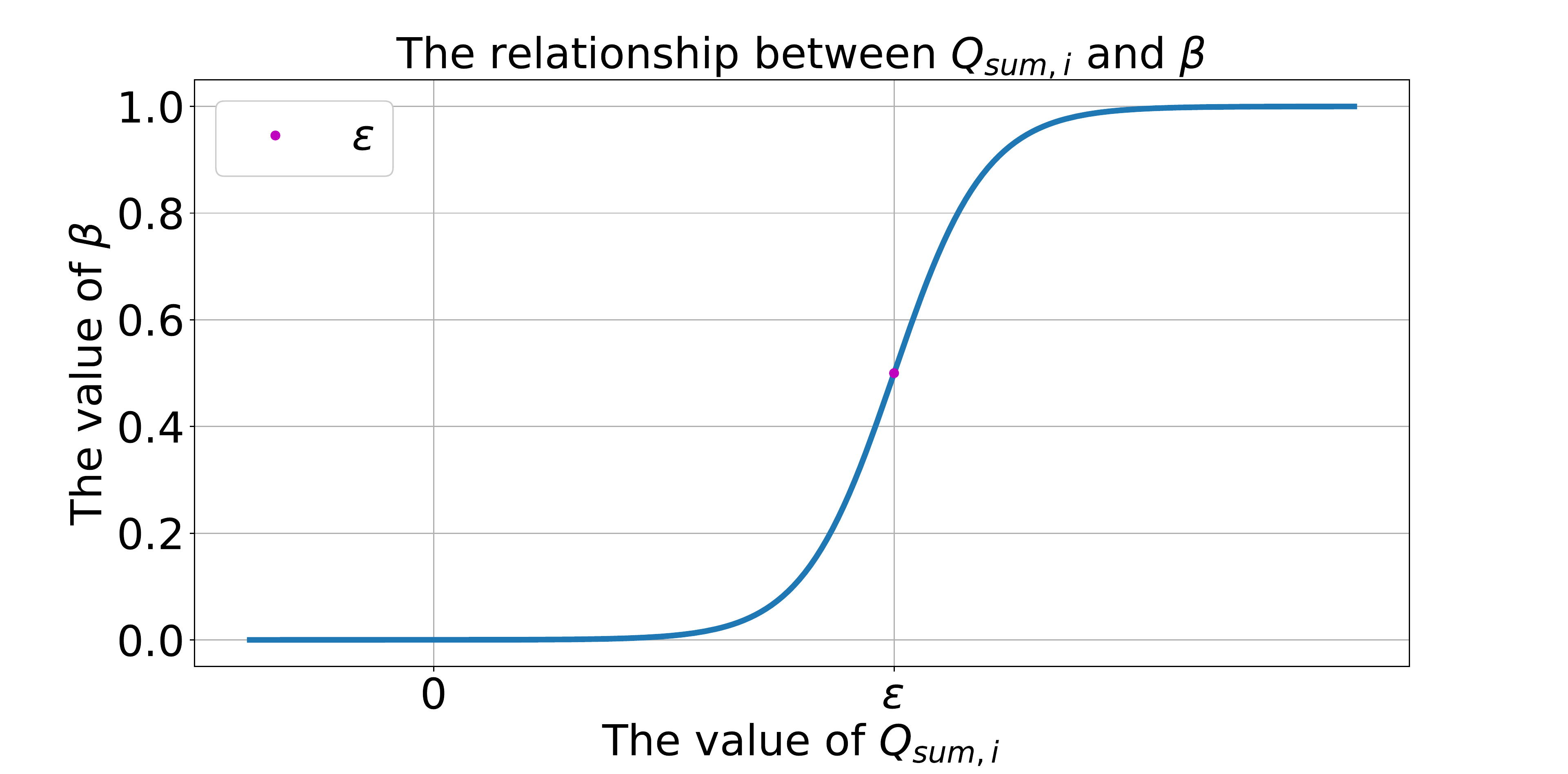}
\caption{The relationship between $Q_{sum,i}$ and $\beta$.} \label{q_beta}
\end{centering}
\end{figure}

\begin{algorithm}[!t]
\small{
\caption{Distributed joint spectrum and power allocation at $d_i$}\label{algr}\centering
\begin{algorithmic}[1]
\STATE Initialize the structure and the parameters of $d_i$;
\STATE Initialize the structure and the parameters of NN from MBS;
\WHILE{$d_i$ is transmitting data}
\STATE Estimate the number of WiFi users with EKF based method;
\STATE Estimate the traffic load $\mathcal{L^U}$ on unlicensed spectrum based on \eqref{luk};
\STATE Normalize the input of NN;
\STATE The prices of all unlicensed channels are calculated based on \eqref{NN};
\STATE Problem(\ref{op2}) is solved to get the resources allocation scheme of $d_i$;
\STATE Calculate the loss value on the basis of \eqref{loss};
\STATE ADD the absolute loss value to $Q_{sum,i}$;
\STATE $d_i$ trains the NN based on gradient descent algorithm in \eqref{NN_TRAIN};
\IF{$T$ can divide $T_{fl}$ evenly}
\STATE MBS collects data and calculates $\hat \phi$ based on \eqref{av_phi};
\STATE $d_i$ updates $\phi_i$ based on \eqref{av_phi_phi_i};
\STATE Reset $Q_{sum,i}$ to $0$;
\ENDIF
\ENDWHILE
\end{algorithmic}}
\end{algorithm}

\section{Numerical Results}\label{s5}

In this section, the simulation results are demonstrated to verify the performance of the proposed distributed D2D-U communication scheme.
In the simulation setup, the relevant parameters of NN and federated learning are demonstrated in Table~\ref{tab1} and the parameters related to the D2D-U and WiFi networks are given in Table~\ref{tab2}.
Since the real-time performance of the algorithm is required in practice, we use shallow fully connected neural networks to reduce algorithm complexity and for quickly convergence during online training processing.
Then during an iteration time slot, D2D-U link will first estimate WiFi traffic load based on EKF method mentioned before, then the price of the unlicensed channels can be derived based on the proposed method. With the unlicensed channels' price as well as WiFi traffic load, D2D-U link can decide its spectrum and power allocation scheme. Finaly NN is trained based on the transmission transmission of D2D-U link and the little signaling interaction with MBS.

\begin{table}
\caption{Parameters of NN.}\label{tab1}
\begin{tabular}{lp{3cm}}
\hline\noalign{\smallskip}
Parameters & Value\\
\noalign{\smallskip}\hline\noalign{\smallskip}
Learning rate $\alpha$ &  $0.0001$\\

Number of hidden layers & $2$\\

Number of neurons & $32$/$32$\\

Active function & $\tanh$/$\tanh$/$\tanh$/$w\times \rm{sigmoid}$\\

Max value of output $w$ & $10$\\

Fairness-based loss $q$ & $0.01$\\

Conflict multiplication loss $v_1$ & $0.03$\\

Maximum throughput loss $v_2$ & $-0.03$\\

Federated learning param $\gamma$ & $1.2$\\

Federated learning param $\epsilon$ & $0.4$\\

Federated learning period $T_{fl}$ & $100$\\
\noalign{\smallskip}\hline
\end{tabular}

\end{table}

\begin{table}\center
\caption{Parameters of D2D-U links.}\label{tab2}
\begin{tabular}{ll}
\hline\noalign{\smallskip}
Parameters & Value\\
\noalign{\smallskip}\hline\noalign{\smallskip}
Total power control $p_c$ &  $35{\rm dBm}$\\

Power control on one unlicensed channel $p_u$& $23{\rm dBm}$\\

AWGN noise power $N$ & $-95{\rm dBm}$\\

Total assets $C$ & $1$\\
\noalign{\smallskip}\hline
\end{tabular}

\end{table}

\subsection{Effectiveness on the proposed NN}
\begin{figure}
\begin{centering}
\includegraphics[width=0.5\textwidth]{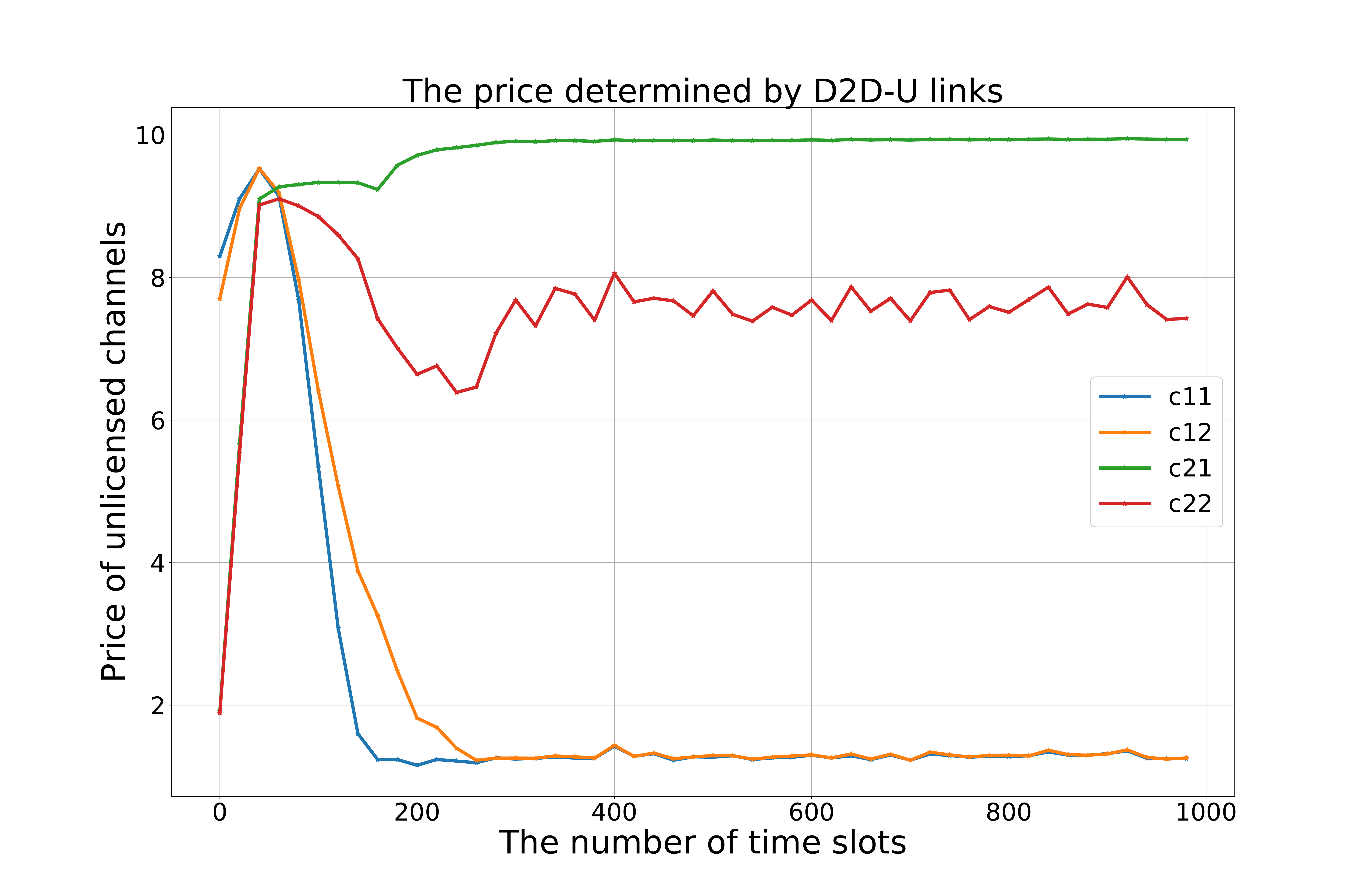}
\caption{The price determined by D2D-U links.} \label{simulation1}
\end{centering}
\end{figure}

We first verify the effectiveness on the proposed scheme, assuming that there are two D2D-U links, $d_1~{\rm{and}}~d_2$, and two independent unlicensed channels, $u_1~{\rm{and}}~u_2$. The traffic load of $d_1$ is set to be larger than $d_2$ and the WiFi
traffic load on unlicensed channel $u_1$ is set to be less than that on $u_2$.
Then the output prices from the NNs of two D2D-U links on two unlicensed channels are illustrated in Fig.~\ref{simulation1}.
It can be observed that after about $250$ iterations, the D2D-U system has achieved convergence, which means that the loss value of NN is tiny and each D2D-U link can stably estimate the price of unlicensed channels.

\begin{figure}
\begin{centering}
\includegraphics[width=0.5\textwidth]{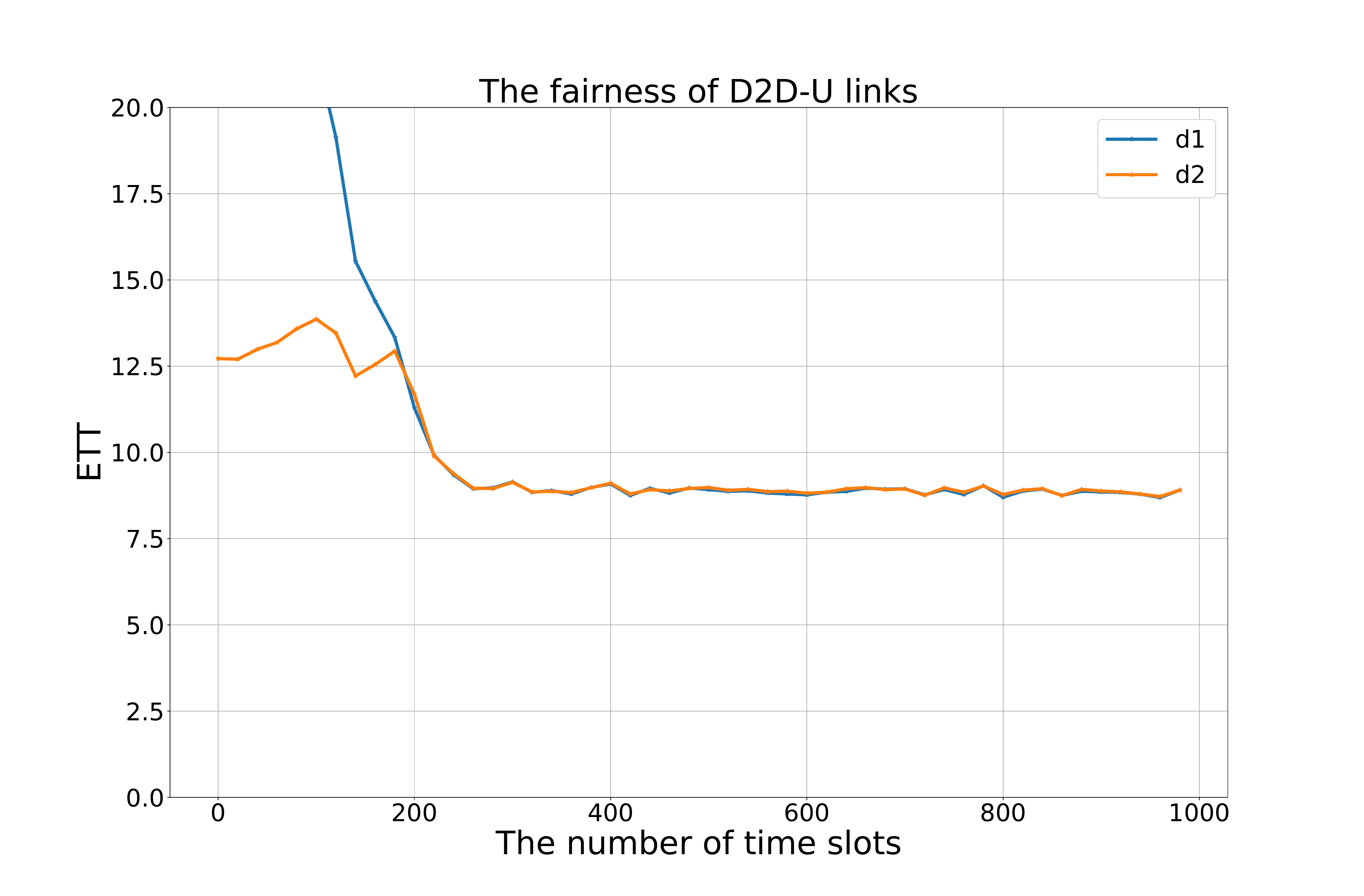}
\caption{The transmission fairness among D2D links.} \label{simulation2}
\end{centering}
\end{figure}

Here we can find that the prices of $u_1$ and $u_2$ is much more cheaper for $d_1$  with heavy transmission tasks, which implies that $d_1$ is encouraged to use more unlicensed spectrum resources.
In addition, since the WiFi traffic load on $u_1$ is lighter than that on $u_2$, the value of $c_{1,1}$ is smaller than $c_{1,2}$ and $d_1$ will select the unlicensed channel $u_1$ in the first place.
As for $d_2$, since $u_1$ has priority to be selected by $d_1$ with much less price, to avoid transmission collision, $c_{2,1}$ is trained larger than $c_{2,2}$ and $d_2$ will mainly use $u_2$ for transmission. Fig.~\ref{simulation2} shows the fairness between $d_1$ and $d_2$, where we can observe that as the prices converge, $\frac{l^D_1}{R_1}$ is equal to $\frac{l^D_2}{R_2}$ and the fairness between $d_1$ and $d_2$ is achieved.

\subsection{Verification on the fairness}
To further illustrate the fairness of the proposed algorithm, the performance of D2D-U system with more D2D-U links and unlicensed channels is analyzed and the result is compared with centralized algorithm which concentrates on maximizing system throughput.
In the centralized algorithm, price based model is not applied and all the parameters are collected at MBS. A joint optimization problem is then built and solved to maximize the throughput of D2D-U system while guaranteeing the harmonious coexistence with the WiFi
networks.
The ETT value comparison of two schemes is shown in Table~\ref{tab3}. When the normalized traffic load of D2D-U links is `1', the actual traffic load is $10e8{\rm bits}$.
It can be found that the proposed price based distributed method can adaptively allocate resources at the D2D-U pairs with respect to the traffic load and channel conditions to guarantee the fairness. On the other hand, the centralized algorithm is far from achieving the fairness.

\begin{figure}
\begin{centering}
\includegraphics[width=0.5\textwidth]{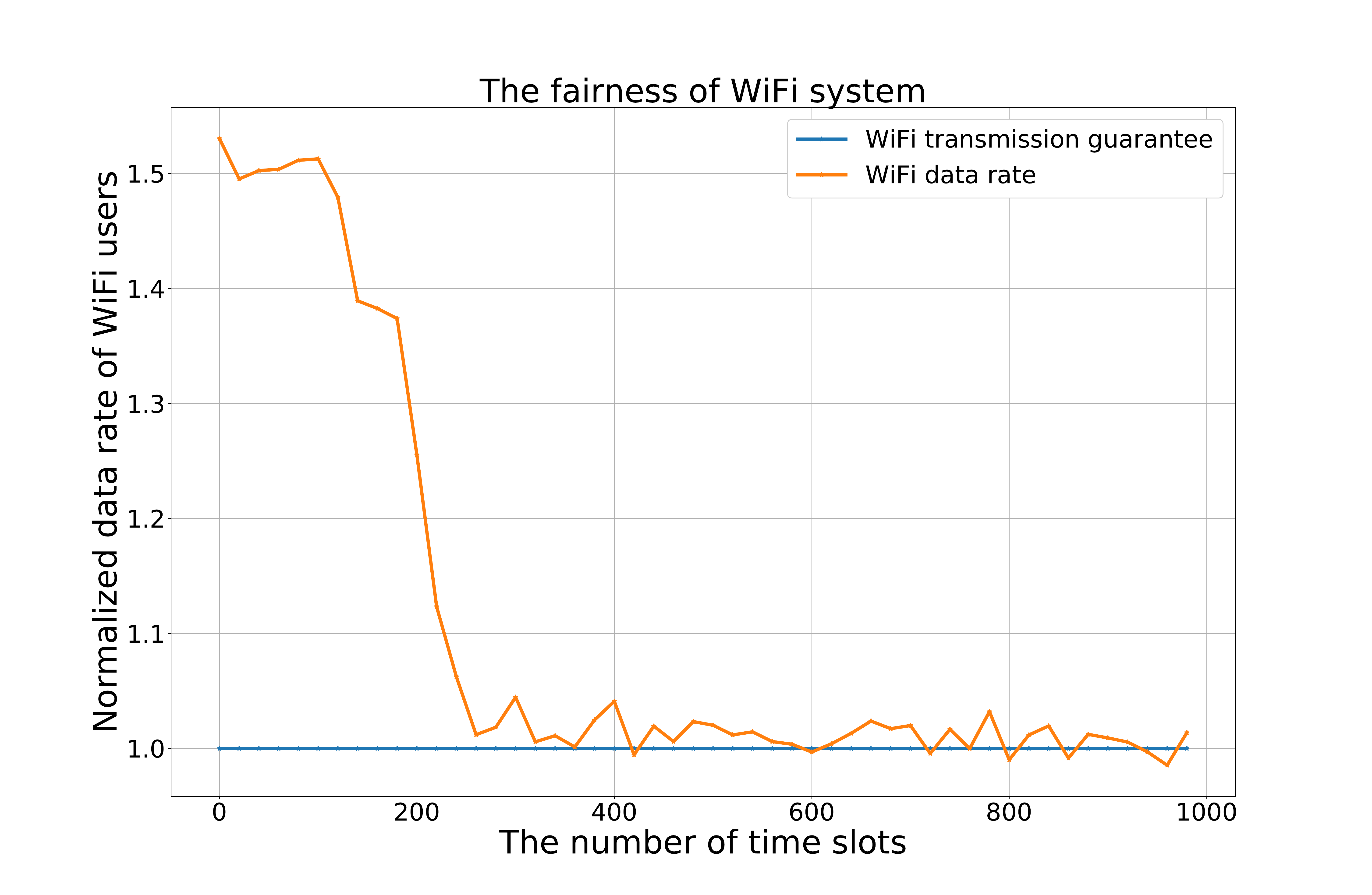}
\caption{Achieved fairness of WiFi system.} \label{simulation3}
\end{centering}

\end{figure}

The fair coexistence with WiFi networks by the proposed scheme is depicted in Fig.~\ref{simulation3}, where the actual WiFi traffic load is normalized based on the the basic WiFi throughput guarantee calculated in \eqref{rk2}. Simulation result shows that after the convergence of the price, the WiFi throughput is basically equal to the basic WiFi throughput guarantee. Due to the set of $v_2$ which encourages D2D-U links to use more spectrum resources with less collision, there are a little transmission collision which leads to the tiny impairment to the WiFi system throughput.

\begin{table}\center
\caption{The comparison of transmission fairness.}\label{tab3}

\begin{tabular}{p{1.6cm}p{1.6cm}p{1.6cm}p{1.6cm}}
\hline\noalign{\smallskip}
Users number & Normalized traffic load & ETT(price-based) & ETT(max throughput) \\
\noalign{\smallskip}\hline\noalign{\smallskip}
$1$ & $0.8$ &8.779 &14.010\\

$2$ & $0.6$ & 8.788 & 10.508\\

$3$ & $0.4$ & 8.750 & 7.005\\

$4$ & $0.2$ & 8.810 & 3.503\\
\noalign{\smallskip}\hline
\end{tabular}

\end{table}

\subsection{Effectiveness of federated learning }
\begin{figure}
\begin{centering}
\includegraphics[width=0.5\textwidth]{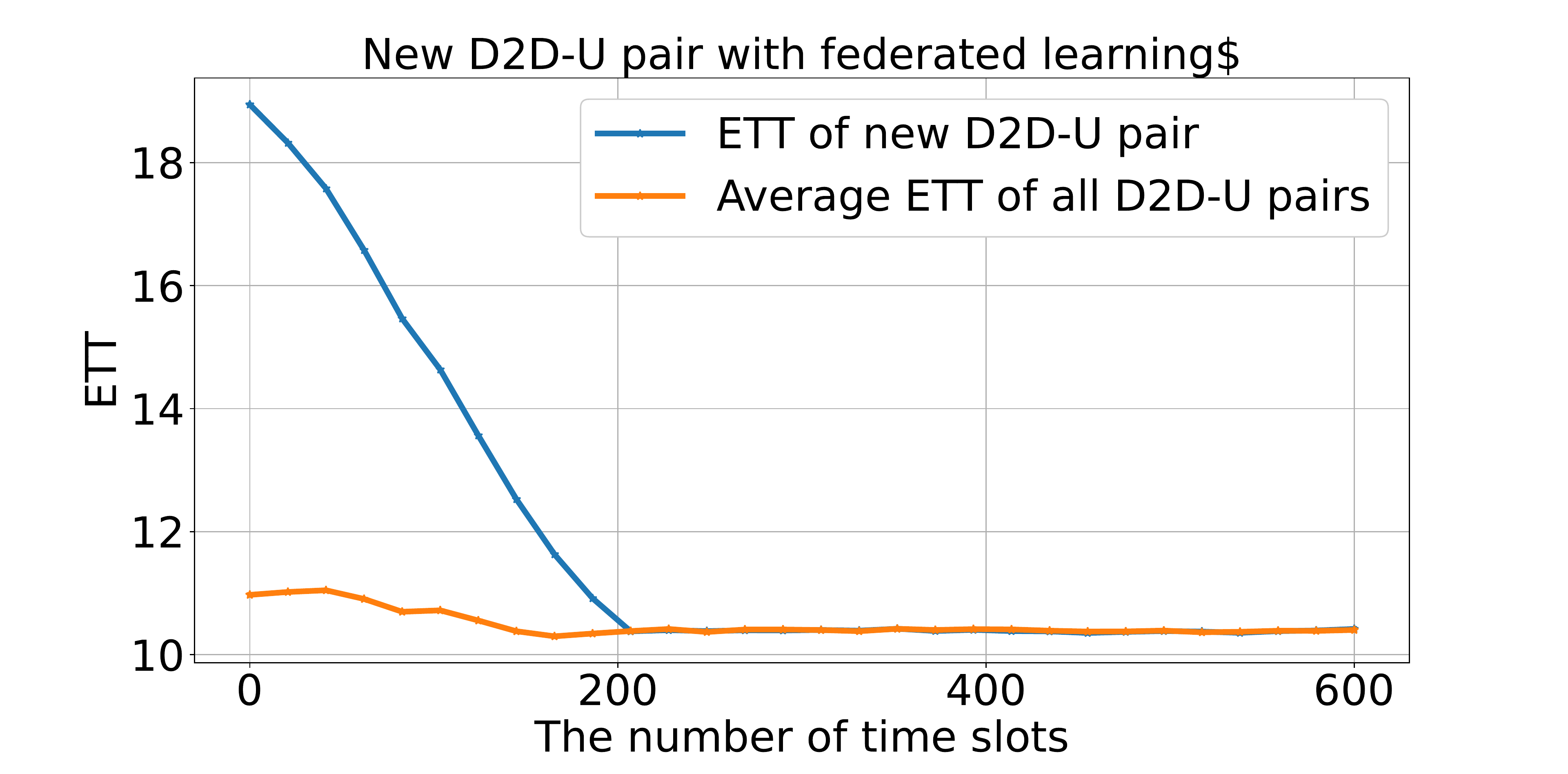}
\caption{The ETT of new D2D-U pair with federated learning.} \label{fd1}
\end{centering}
\end{figure}

\begin{figure}
\begin{centering}
\includegraphics[width=0.5\textwidth]{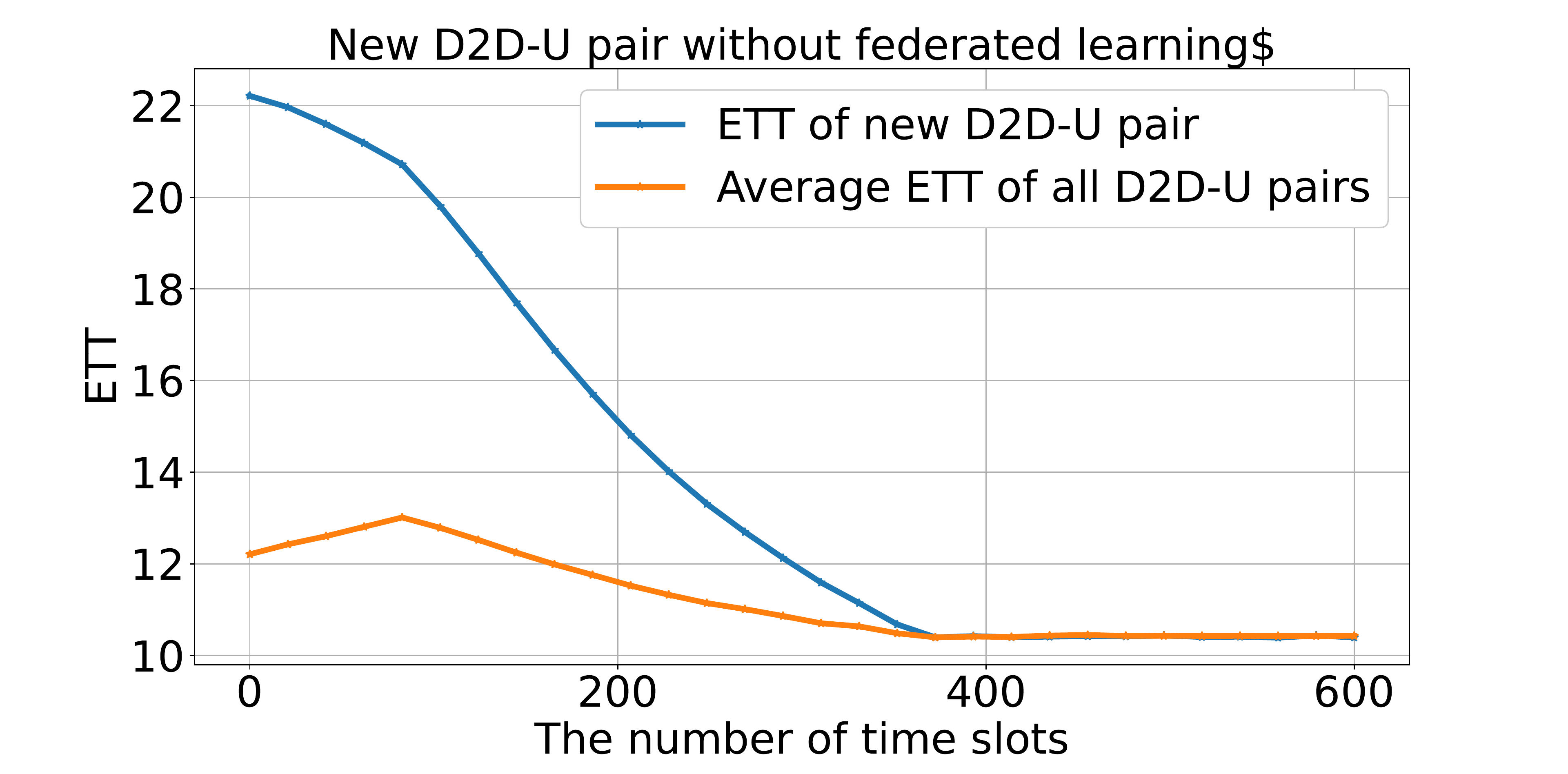}
\caption{The ETT of new D2D-U pair without federated learning.} \label{fd2}
\end{centering}
\end{figure}

To further verify the effectiveness of federated learning, the convergence results of new D2D-U pairs with different NN parameters are illustrate in Fig.~\ref{fd1} and Fig.~\ref{fd2}, respectively. Here the new D2D-U link in Fig.~\ref{fd1} and Fig.~\ref{fd2} are the same instead of NN parameters and system has converged before new users coming, then ETT value of new D2D-pairs and the average ETT value of all D2D-U pairs are provided to prove the performance of the proposed scheme. In Fig.~\ref{fd2}, NN parameters are initialized randomly where new D2D-U pair spends more time to achieve the optimal output and the fluctuation of system performance is also greater. On the contrary, with the federated learning parameters saved in MBS, new D2D-pair in Fig.~\ref{fd1} completes convergence faster and the system performance is more stable.

\subsection{Achievable data rates}

\begin{figure}
\begin{centering}
\includegraphics[width=0.5\textwidth]{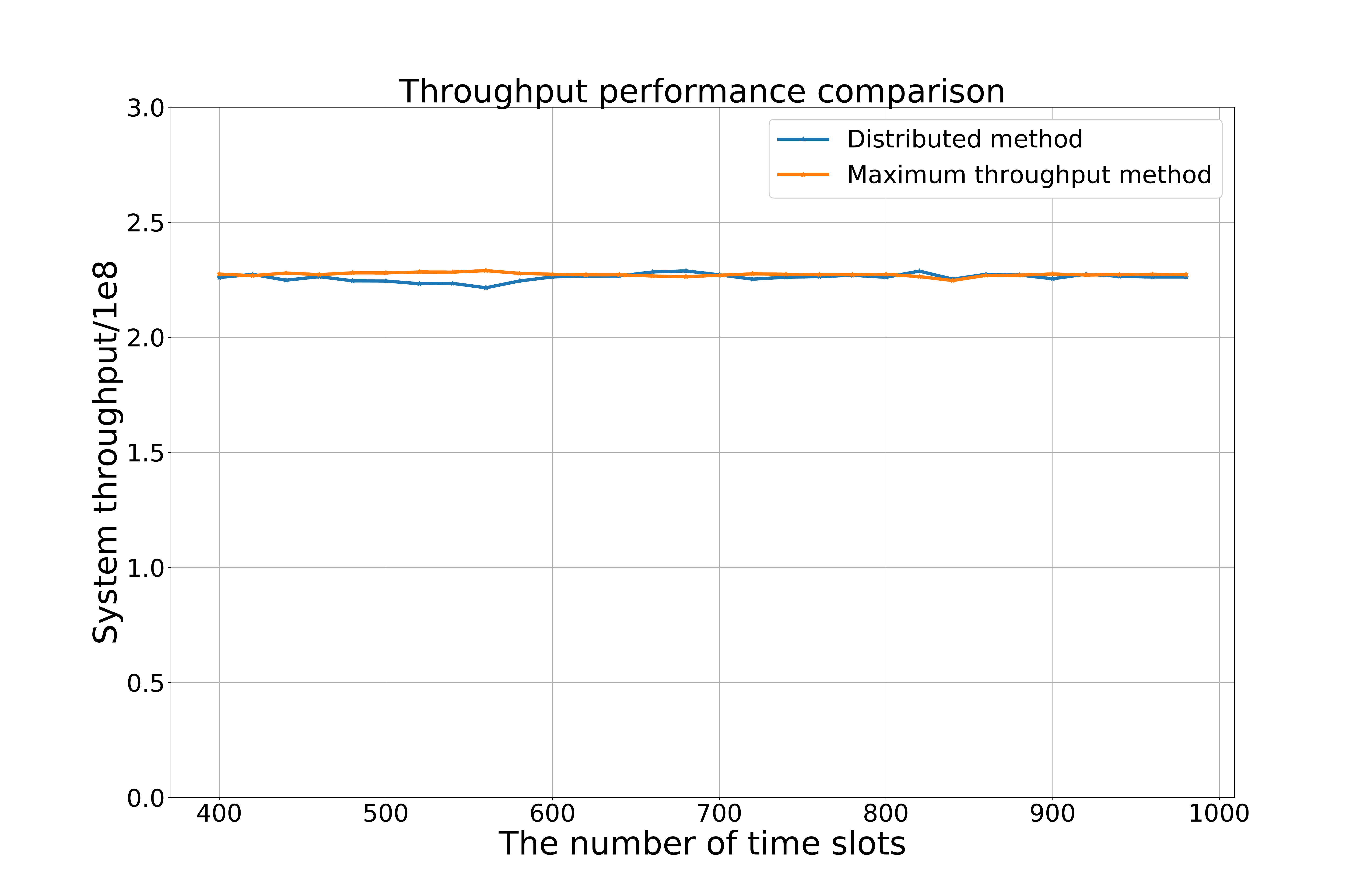}
\caption{The comparison of system throughput between price based method and maximum throughput method.} \label{simulation4}
\end{centering}
\end{figure}

\begin{figure}

\includegraphics[width=0.53\textwidth]{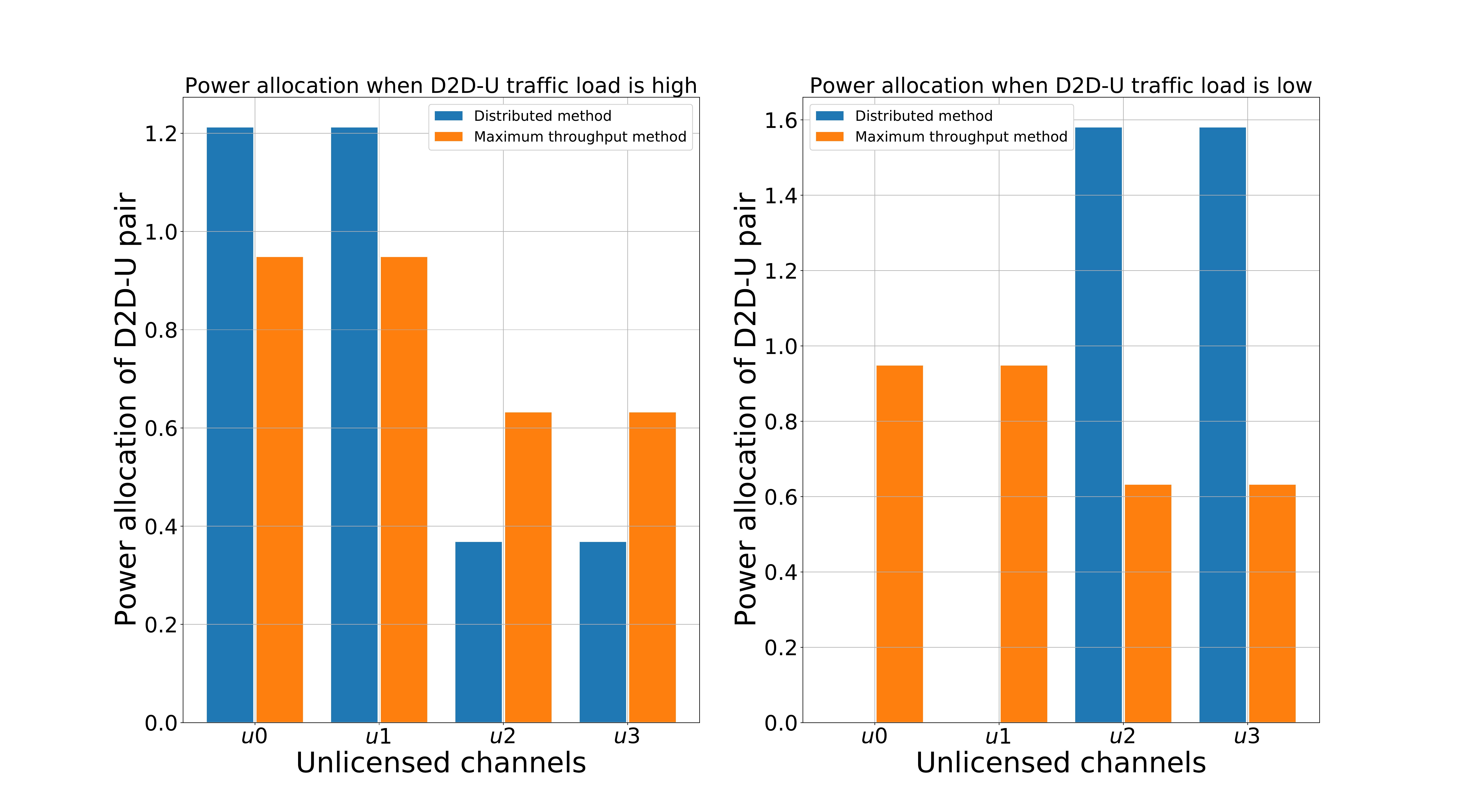}
\caption{The comparison of power allocation between price based method and maximum throughput method.} \label{simulation5}

\end{figure}

Fig.~\ref{simulation4} demonstrates the total achievable data rates of D2D-U links with centralized and proposed scheme, respectively.
From the figure, we can observe that the data rates obtained by the proposed are close to that by the centralized scheme, which indicates that the proposed method can almost maximize the system performance while ensuring the fairness of D2D-U pairs. Besides, the details of D2D-U link power allocation is illustrated in Fig.~\ref{simulation5}, where the unlicensed channel traffic load on $u_0$ and $u_1$ is low and on $u_2$ and $u_3$ is high. The abscissa of Fig.~\ref{simulation5} is different unlicensed channels and the ordinate is $\eta_{i,j}$ of related D2D-U pairs and unlicensed channels.
It can be observed that in the price based model, the D2D-U link with more traffic load reuses the spectrum resources of more ideal channels and the D2D-U link with less traffic load chooses to reuse more crowded channels. While in the centralized method, the change of D2D-U traffic load has no effect on its power allocation scheme.
Therefore, simulation results justify that the proposed method can guarantee the fairness among D2D-U pairs with least lose on the data rates comparing with the centralized optimal solution.

\section{Conclusion}\label{s6}

In this paper, in order to reuse the spectrum resources on the unlicensed spectrum to improve the performance of D2D-U system, a distributed power and spectrum allocation mechanism with adaptive price adjustment scheme is proposed.
An unsupervised online learning structure is employed on each D2D-U link to estimate the prices of all perceived unlicensed channels, a federated learning based approach is utilized to improve system stability and performance.
Then the power and spectrum optimization models can be formulated and solved by D2D-U pairs to access the unlicensed spectrum.
Numerical simulation proves that the proposed algorithm allows D2D-U link to maximize data-rate while ensuring the fairness of WiFi system and the fairness among different D2D-U users.

\begin{acknowledgements}
This work was supported in part by the National Natural Science Foundation of China under Grant No. 61771429, No. 61703368, in part by Zhejiang University City College Scientific Research Foundation under Grant No. JZD18002, in part by Zhejiang Provincial Key Laboratory of Information Processing, Communication and Networking, and in part by the selective Grants for Postdoctoral Programs ZJ2020035 in Zhejiang Province, in part by ROIS NII Open Collaborative Research 2020-20S0502, and JSPS KAKENHI grant numbers 18KK0279, 19H04093 and 20H00592.
\end{acknowledgements}

%
%



\end{document}